\documentclass[3p,times]{elsarticle}

\usepackage{ecrc}

\usepackage{epstopdf}

\volume{00}

\firstpage{1}

\journalname{Nuclear Physics A}

\runauth{A. Festanti for the ALICE Collaboration}

\jid{nupha}

\jnltitlelogo{Nuclear Physics A}

\usepackage{graphicx}
\usepackage{amsmath,amssymb}
\usepackage{subfig}
\usepackage{multirow}
\usepackage{array}

\usepackage{lineno}

\biboptions{sort&compress}

\begin{document}
\begin{frontmatter}



\title{Heavy-flavour production and nuclear modification factor\\
in Pb--Pb collisions at $\sqrt{s_{\rm NN}}=2.76~{\rm TeV}$ with ALICE}

\author{Andrea Festanti (for the ALICE Collaboration)}
\address{Universit\`a degli Studi di Padova, INFN - Sez. di Padova,
  via Marzolo 8, 35131 Padova, Italy}




\begin{abstract}
The measurement of heavy-flavour production and nuclear modification
factor in heavy-ion collisions provides insights into the mechanisms of parton energy loss in the hot and dense medium formed
in these collisions. ALICE results on
heavy-flavour decay electrons and muons and on D mesons are presented.
\end{abstract}

\begin{keyword}
Heavy-ion collisions \sep QGP \sep Heavy flavour \sep Nuclear
modification factor \sep ALICE

\end{keyword}

\end{frontmatter}



\section{Introduction}
\label{intro}

Charm and beauty quarks are effective probes of the properties
of the hot and dense strongly-interacting medium formed in high-energy nuclear
collision, the Quark-Gluon Plasma (QGP). They are sensitive to
the transport properties of the medium, since they are predominantly produced
on a short time scale in primary partonic
scattering processes with large virtuality $Q^2$. Due to their large masses
($m_{\rm c}\sim 1.5~{\rm GeV}/c^2$ and $m_{\rm b}\sim 4.5~{\rm GeV}/c^2$) and the minimum
virtuality, $Q_{\rm min}^2=(2m_{\rm Q})^2$, required for the production of a $\rm {Q\overline{Q}}$ pair, the temporal and
spatial scales, $\Delta\tau\sim\Delta r\sim 1/Q$, are sufficiently small for the production to be
unaffected by the QGP medium.

For hard processes, in the absence of nuclear
and medium effects, a nucleus--nucleus (A--A) or a proton--nucleus (p--A)
collision would behave as a superposition of independent
nucleon--nucleon (NN) collisions. The charm and beauty differential
yields ${\rm d}N/{\rm d}p_{\rm T}$ would scale from pp to A--A or p--A proportionally to the
number $N_{\rm coll}$ of inelastic NN collisions (binary scaling). The
binary scaling is ``broken'' by initial-state and final-state effects. In
the initial state, the nuclear environment affects the quark and gluon
distributions, which are modified in bound nucleons depending on the
parton fractional momentum $x$ and the atomic mass number A
\cite{ref1}. Partons can also lose energy in the initial stages of the
collision via initial state radiation \cite{ref2} or experience transverse
momentum broadening due to multiple soft collisions before the ${\rm Q\overline{Q}}$ pair is produced \cite{ref3}. The final-state
effects are related to the heavy-quark
interaction with the medium. The most relevant
effect is partonic energy loss due to medium-induced gluon radiation
(inelastic processes) \cite{ref4, ref5} and collisions with medium constituents (elastic
processes) \cite{ref6}. The modification of binary scaling is quantified via the nuclear
modification factor, defined as $R_{\rm AA}(p_{\rm T})=\frac{({\rm d}N/{\rm
    d}p_{\rm T})_{\rm AA}}{\langle T_{\rm AA} \rangle
  ({\rm d}\sigma/{\rm d}p_{\rm T})_{\rm pp}}$, where $\langle T_{\rm
  AA} \rangle$ is the average nuclear overlap function calculated with
the Glauber model in the considered centrality range. Quarks are predicted to
lose less energy than gluons due to their smaller colour coupling
factor. In addition, the dead-cone effect is expected to
reduce small-angle gluon radiation, thus the energy loss, for heavy quarks with respect to light
quarks \cite{ref7, ref8}. This effect can be tested by comparing the $R_{\rm
  AA}$ suppression of the mostly gluon-originated
light-flavour hadrons and that of D and B mesons. However, the comparison of heavy-flavour hadron and pion $R_{\rm  AA}$ cannot
be interpreted directly as a comparison of charm, beauty and gluon
energy losses due to the different fragmentation functions and slope
of the $p_{\rm T}$-differential cross sections \cite{ref9, ref10}.

\vspace{-3.5mm}
\section{Analysis and results}
\label{anRes}
Charm and beauty production was measured with ALICE
\cite{ref11, ref12} in Pb--Pb collisions at $\sqrt{s_{\rm
    NN}}=2.76~{\rm TeV}$ 
using electrons and muons from semi-leptonic decays of
heavy-flavour hadrons and fully reconstructed D-meson hadronic decays.
Electron tracks were identified in the central rapidity region using
the specific energy loss in the gas of the Time Projection Chamber (TPC)
combined with the information provided by the Time Of Flight (TOF)
detector or the
elecromagnetic calorimeter (EMCAL). The data were collected
using minimum bias and EMCAL triggers. Muon tracks were reconstructed in minimum bias trigger events in the Forward Muon Spectrometer
($-4<\eta<-2.5$). D mesons were reconstructed at mid-rapidity ($|y|<0.5$) in minimum
bias Pb--Pb collisions 
via their hadronic decay channels: ${\rm D^0\to K^-\pi^+}$ (with branching ratio, BR, of $3.88 \pm 0.05\%$), ${\rm
  D^+\to K^-\pi^+\pi^+}$ (BR of $9.13 \pm 0.19\%$), ${\rm D^{*+}\to
  D^0\pi^+}$ (BR of $67.7\pm
0.5\%$) and ${\rm
  D_s^{+}\to \phi\pi^+\to K^-K^+\pi^+}$ (BR of $2.28 \pm 0.12\%$) \cite{ref13} and
their charge conjugates. D-meson selection was based on the reconstruction
of decay vertices displaced by a few hundred $\mu$m from the interaction vertex, exploiting the
the high track spatial resolution close to the
interaction vertex of the collision, granted by the Inner Tracking
System detector (ITS). Charged pions and kaons were identified using
TPC and TOF signals.

The reference pp cross section of heavy-flavour decay electrons and D mesons at $\sqrt{s_{\rm NN}}=2.76~{\rm TeV}$
was obtained by a pQCD-based energy scaling of the $p_{\rm
  T}$-differential cross section measured at $\sqrt{s}=7~{\rm
  TeV}$. The scaling factor and its uncertainties were evaluated as
explained in \cite{ref14}. The muon $R_{\rm
  AA}$ was calculated using the pp cross section measured at
$\sqrt{s}=2.76~{\rm TeV}$ \cite{ref15}.

\begin{figure}[!h]
\begin{minipage}{1.\textwidth}
\begin{center}
\begin{tabular}{cc}
\multirow{-2}[20]{*}{\subfloat{\includegraphics*[width=0.38\textwidth]{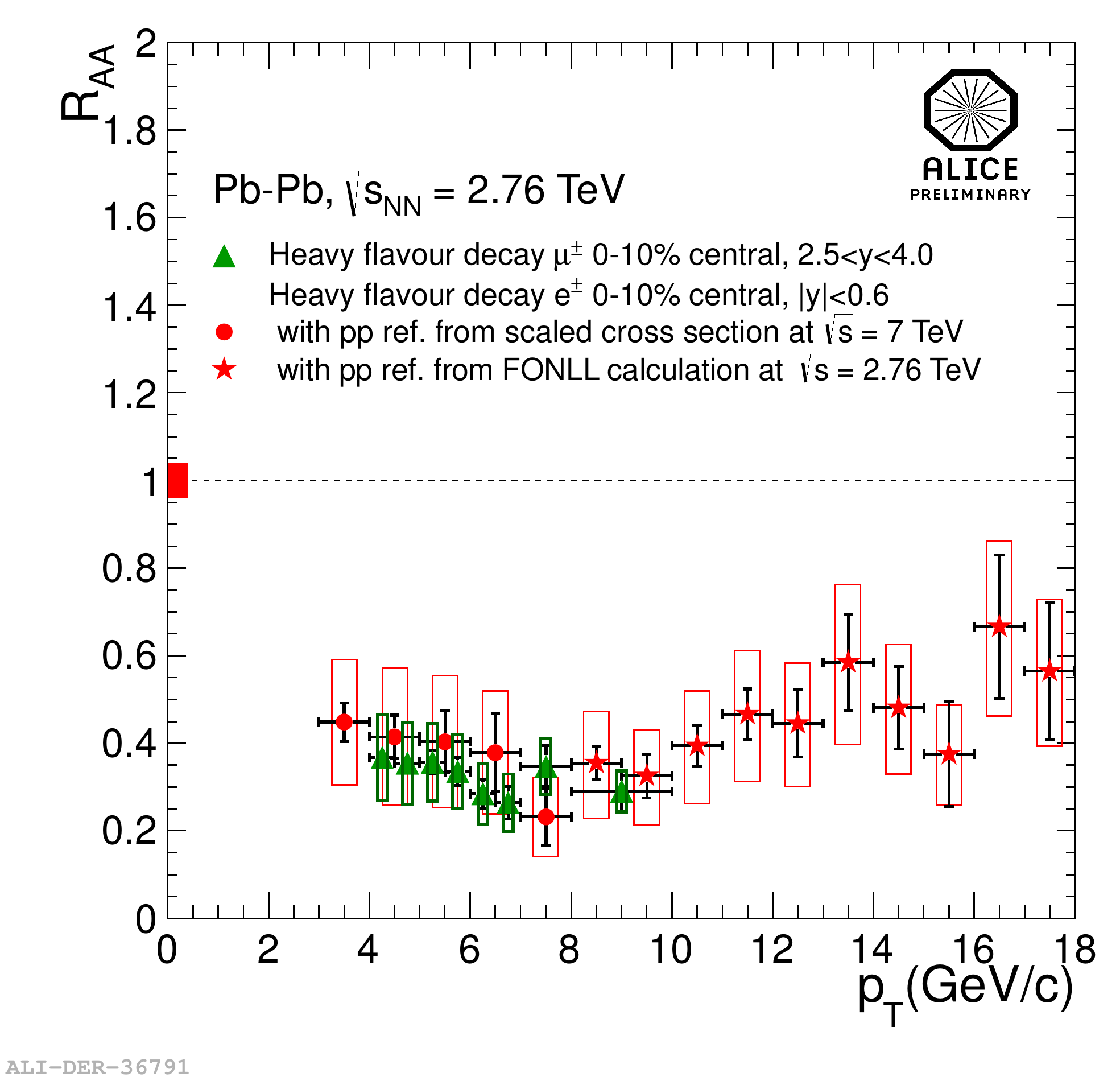}}}
&
\subfloat{\includegraphics*[width=0.35\textwidth]{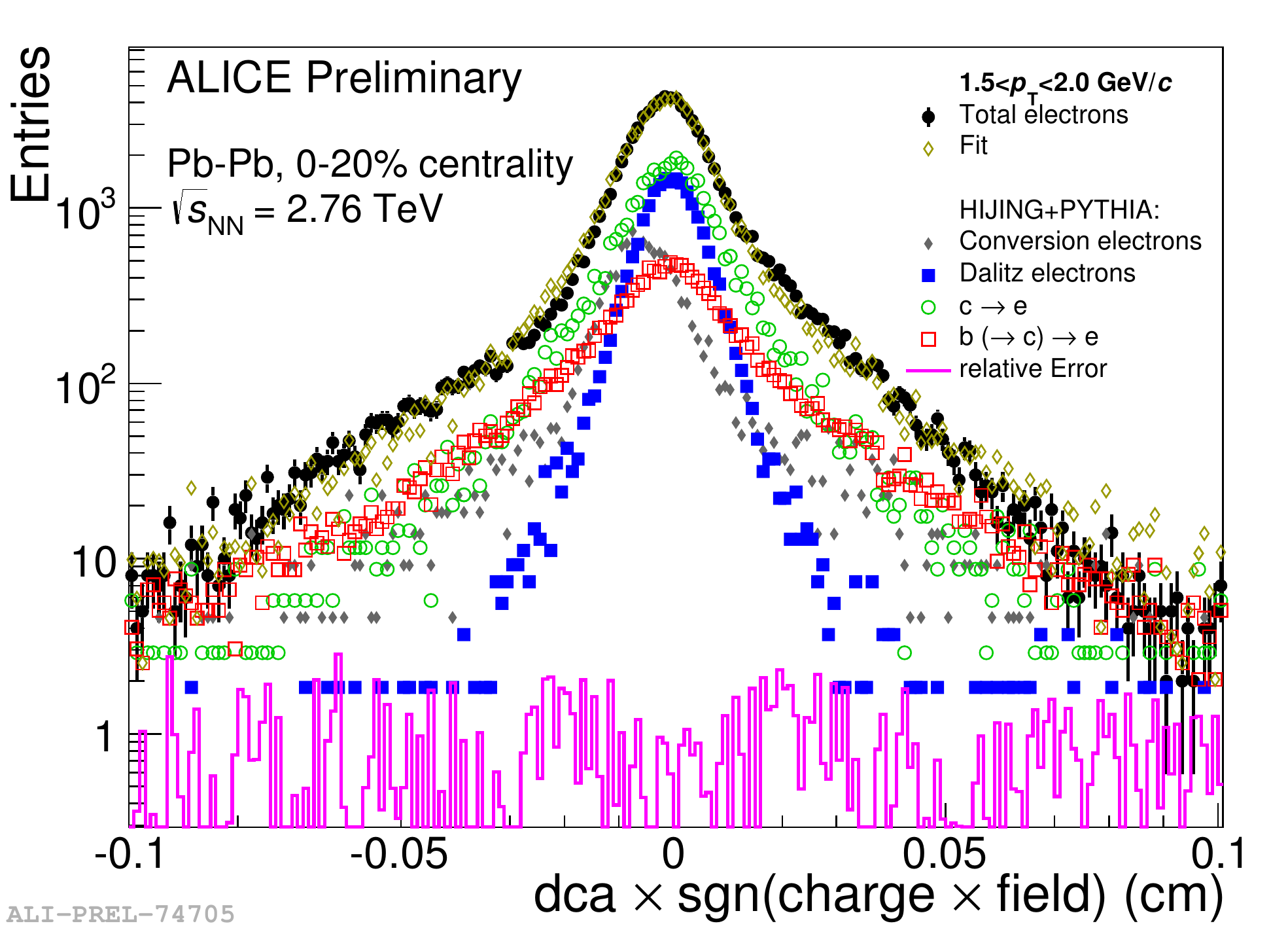}}\\ &
\subfloat{\includegraphics*[width=0.38\textwidth]{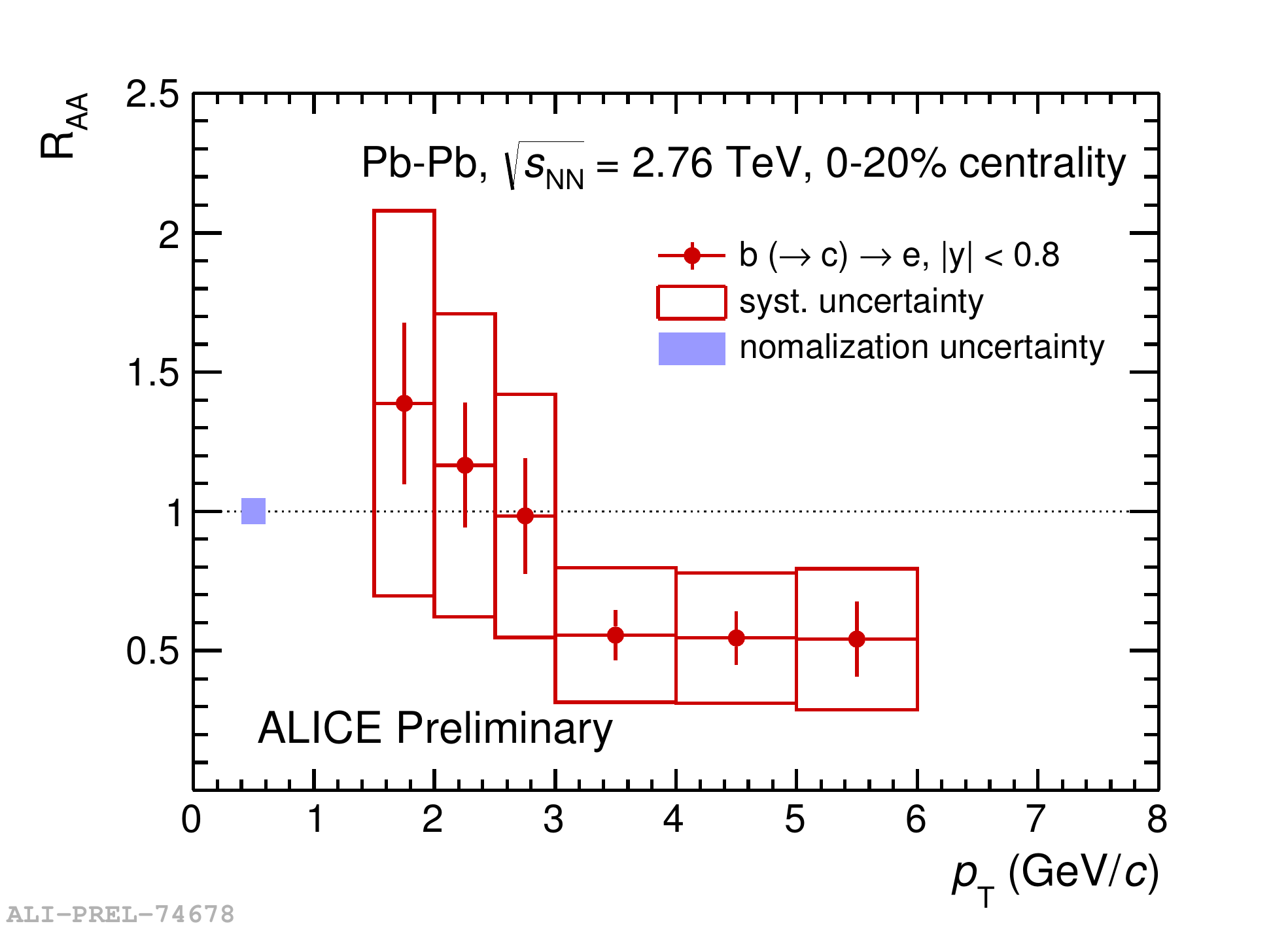}}
\end{tabular}
\caption{Left: $R_{\rm AA}$ of heavy-flavour decay electrons at central
 rapidity and heavy-flavour decay muons at forward rapidity in central
(0--10\%) Pb--Pb collisions, as a function of $p_{\rm T}$. Right top: Inclusive electron impact parameter distribution (black
  points) fitted with templates for the various electron sources
  obtained from simulations. Right bottom: beauty decay electron $R_{\rm AA}$ as a funtion
  of $p_{\rm T}$ measured in central Pb--Pb collisions.}
\label{fig:HFEMRaa010}
\end{center}
\end{minipage}
\end{figure}
The left panel of Fig.~\ref{fig:HFEMRaa010} shows the heavy-flavour
decay electron and the heavy-flavour decay muon $R_{\rm AA}$, measured
in the
10\% most central Pb--Pb collisions, at
central ($|y|<0.6$) and forward rapidity ($2.5<y<4$) respectively. A clear suppression is observed for both
electrons and muons in the measured $p_{\rm T}$ range and it is compatible within uncertainties at central
and forward rapidity. ALICE measured also the nuclear modification factor,
$R_{\rm pPb}$, of heavy-flavour decay electrons and muons in
minimum-bias p--Pb collisions at $\sqrt{s_{\rm NN}}=5.02$ TeV and the results are compatible
with unity \cite{ref16}, indicating that the suppression observed in
central Pb--Pb collisions is due to final state effects.
The fraction of electrons produced by beauty-hadron decays was extracted
from a fit to the electron impact parameter distribution (right top
panel of Fig.~\ref{fig:HFEMRaa010}). The electron sources were included in the fit through templates obtained
from simulations. The data indicate that the beauty-decay
electron $R_{\rm AA}$ is smaller than unity for $p_{\rm T}>3$
GeV/\textit{c} (right bottom panel of Fig.~\ref{fig:HFEMRaa010}). The same
measurement was performed in p--Pb collisions and the compatibility of the
resulting $R_{\rm pPb}$ with unity suggests that in Pb--Pb collisions the b quark is
affected by the interaction with the hot medium.
\begin{figure}[!ht]
\begin{center}
\subfloat{%
  \begin{minipage}[c][1\width]{%
      0.35\textwidth}
    \centering%
    \includegraphics*[width=1\textwidth]{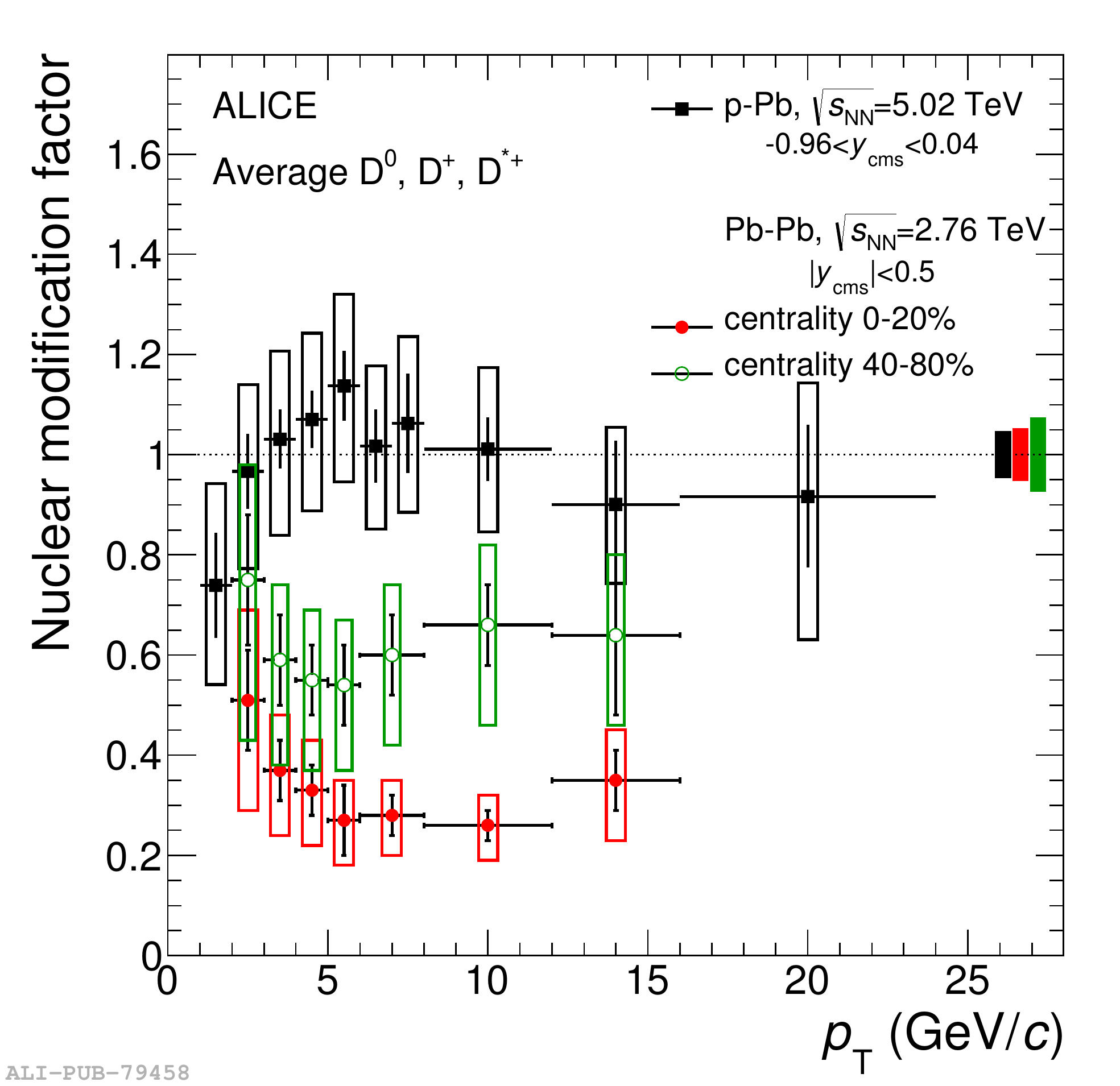}
  \end{minipage}}
\subfloat{%
  \begin{minipage}[c][1\width]{%
      0.34\textwidth}
    \centering%
    \includegraphics*[width=1\textwidth]{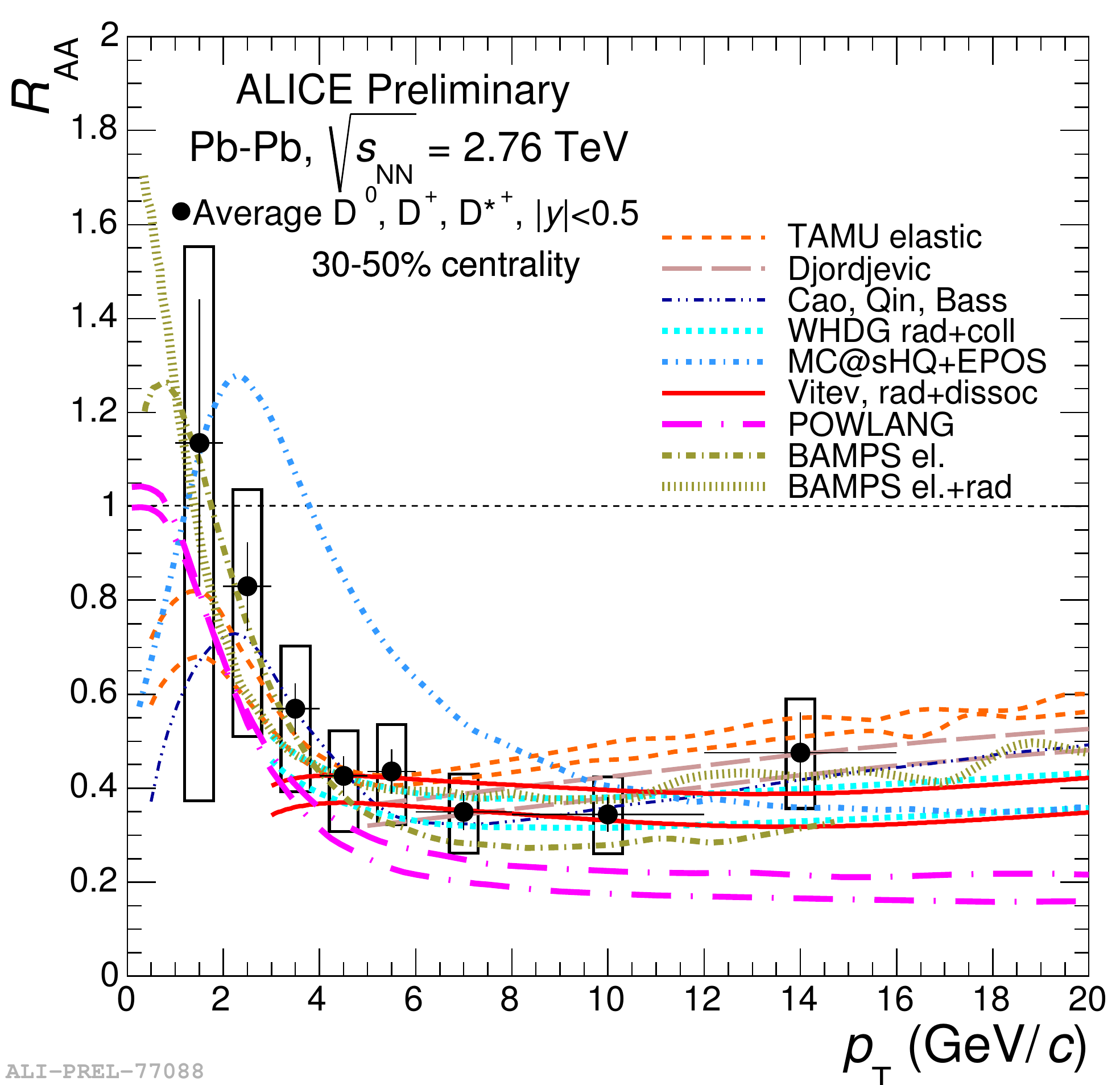}
  \end{minipage}}
\caption{Left: average $R_{\rm pPb}$ of prompt ${\rm D^0}$, ${\rm D^+}$
  and ${\rm D^{*+}}$ mesons as a function of $p_{\rm T}$ compared to
  the prompt D-meson $R_{\rm AA}$ in the 20\% most central Pb--Pb collisons and in
  the 40--80\% centrality class \cite{ref17}. Right: average prompt D-meson $R_{\rm AA}$ in Pb--Pb collisions in the  30--50\% centrality class compared with
  theoretical models including parton energy loss \cite{ref9, ref19}.}
\label{fig:DRaavspt}
\end{center}
\end{figure}

In the left panel of Fig.~\ref{fig:DRaavspt} the comparison of the
average of prompt ${\rm D^0}$, ${\rm D^+}$ and ${\rm D^{*+}}$ nuclear modification factors measured in two centrality
classes of Pb--Pb collisions \cite{ref17} (0--20\% and 40--80\%) and in minimum
bias p--Pb collisions is presented. The suppression observed in the 20\% most central Pb--Pb
collisions (about a factor 4 for $p_{\rm T}>5~{\rm GeV/}c$) is
predominantly induced by final state effects due to charm quark energy
loss in the medium \cite{ref18}. In the right panel of
Fig.~\ref{fig:DRaavspt} the average of prompt ${\rm D^0}$, ${\rm D^+}$
and ${\rm D^{*+}}$ $R_{\rm AA}$ measured in Pb--Pb collisions in the centrality
class 30--50\% is compared with several theoretical models including in-medium
energy loss \cite{ref9, ref19}. A significant suppression is observed also
in the 30--50\% centrality class for $p_{\rm}>4~{\rm GeV}/c$. ALICE
measured also the ${\rm D_{\rm s}^+}$ $R_{\rm AA}$ \cite{ref20},
observing a similar suppression to that of ${\rm D^0}$, ${\rm D^+}$
and ${\rm D^{*+}}$ in the $p_{\rm T}$ interval from 8 to 12
GeV/$c$. At lower $p_{\rm T}$ the current uncertainties does not allow
to draw a clear conclusion. The
comparison of the theoretical predictions with different
observables, such as the D-meson production cross section in 30--50\%,
the $R_{\rm AA}$ in 30--50\%, the $R_{\rm AA}$ in 0--7.5\% and the
azimuthal dependence of the nuclear modification factor in 30--50\%
centrality class \cite{ref21}, allows to constrain the
description of the energy loss mechanisms (see e.g. \cite{ref21} for a
discussion).

\begin{figure}[!h]
\begin{center}
\includegraphics*[width=0.35\textwidth]{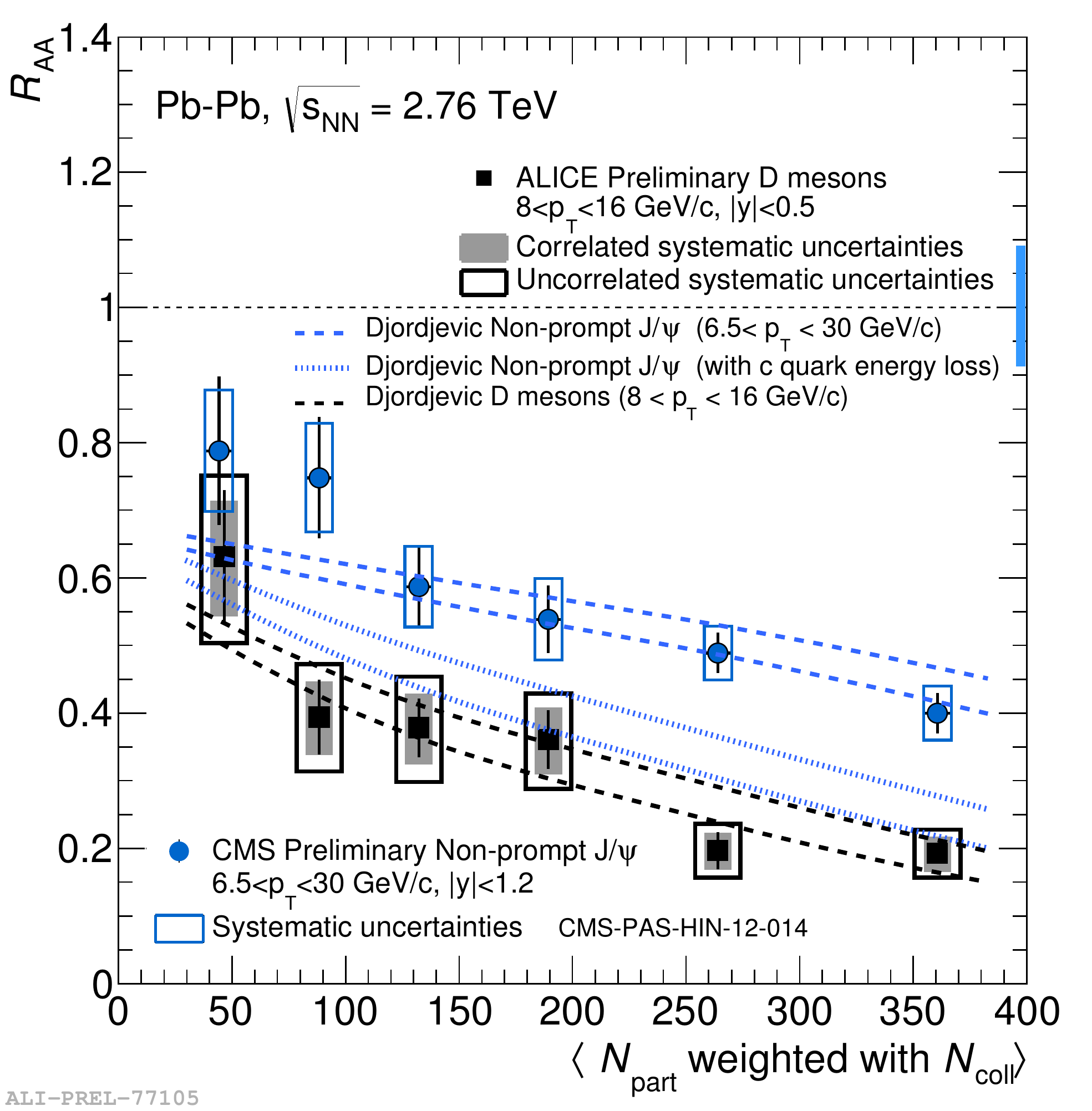}
\includegraphics*[width=0.35\textwidth]{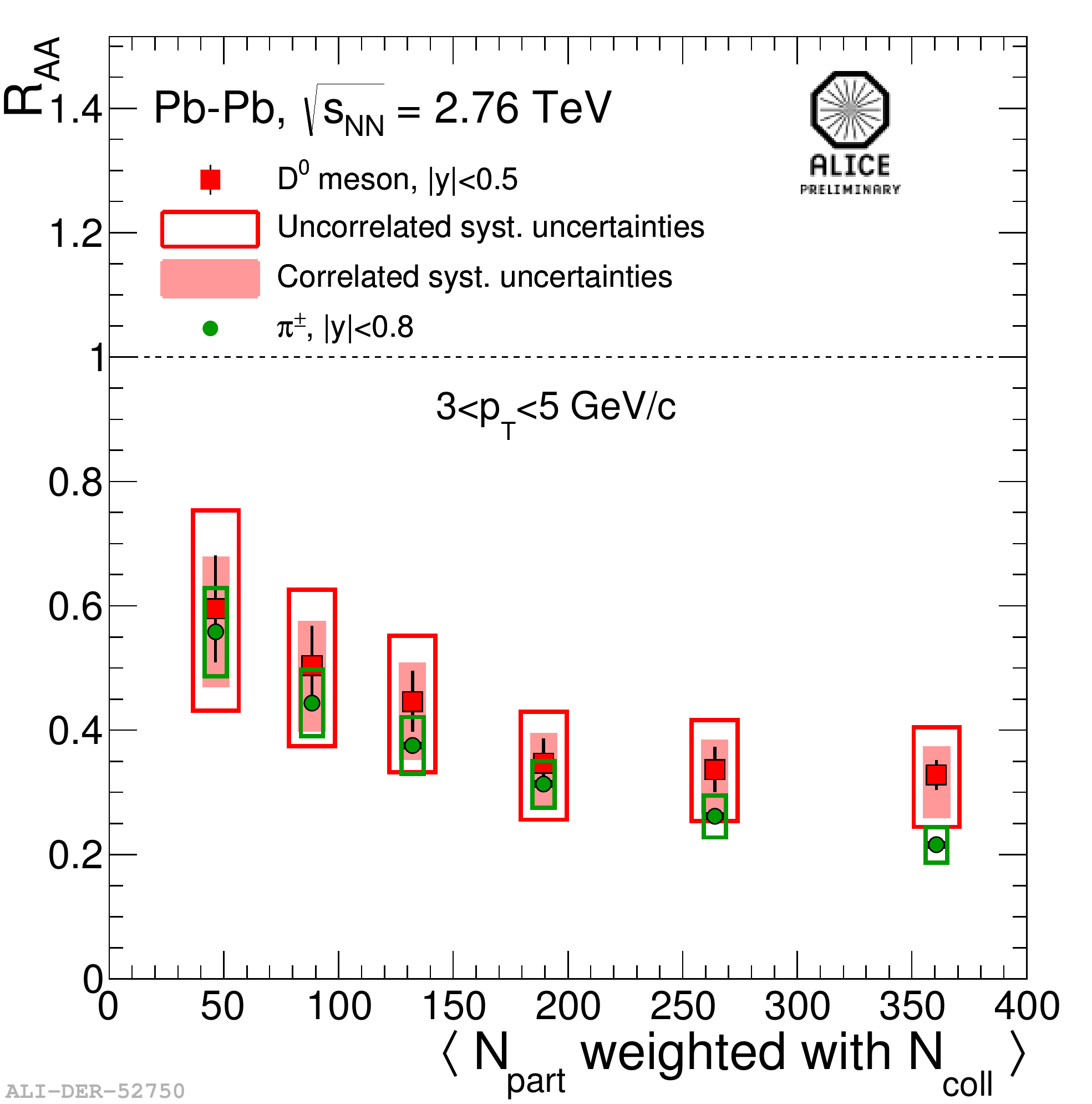}
\caption{Left: centrality dependence of the $R_{\rm AA}$ of prompt D mesons and of J/$\psi$ from B
meson decay (measured by CMS \cite{ref22}) compared with a pQCD model
including mass dependent radiative and collisional energy loss
\cite{ref9}. Right: comparison of ${\rm D^0}$ and charged pion
$R_{\rm AA}$ as a function of centrality for $3<p_{\rm T}<5$ GeV/$c$.}
\label{fig:DRaavsNpart}
\end{center}
\end{figure}
The comparison of the centrality dependence of the $R_{\rm AA}$ of D mesons and of J/$\psi$ from B
meson decays (measured by CMS \cite{ref22}) is displayed in the left panel
of Fig.~\ref{fig:DRaavsNpart}. The
$p_{\rm T}$ range 8--16 GeV/$c$ was chosen for D mesons in order to
have a similar average transverse momentum (about $10~ {\rm GeV}/c$)
than that of B mesons decaying in a ${\rm J}/\psi$ in the measured
$p_{\rm T}$ interval of 6.5--30 GeV/$c$. It shows an indication for a
stronger
suppression for charm than for beauty at high $p_{\rm T}$ in central Pb--Pb collisions. The two measurements are
described by the predictions based on a pQCD model including
mass-dependent radiative and collisional energy loss \cite{ref9}. In
this model the difference in $R_{\rm AA}$ of charm and beauty mesons is mainly
due to the mass dependence of the charm and beauty quark energy
loss, as shown by the curve in which the non-prompt J/$\psi$ $R_{\rm
  AA}$ is calculated assuming that b quarks suffer the same energy
loss as c quarks. The right
panel of Fig.~\ref{fig:DRaavsNpart} shows the comparison of $R_{\rm
  AA}$ as a function of centrality of ${\rm D^0}$ mesons and charged pions for
$3<p_{\rm T}<5~{\rm GeV/}c$. The results
are compatible within uncertainties: better precision is needed to
investigate the expected difference of gluon and light quark energy
loss with respect to charm. 
 
\vspace{-3.5mm}
\section{Conclusions}
\label{concl}
The results obtained with ALICE using the data from the LHC Run-1
(2010-2013) indicate a strong suppression of heavy flavour production in central Pb--Pb collisions for $p_{\rm
  T}>3~{\rm GeV}/c$, observed for heavy-flavour decay electrons and
muons, for electrons from beauty decays and for D mesons. From the comparison with p--Pb measurements, it is
possible to conclude that the suppression observed in Pb--Pb collisions is mainly due to
final state effects, i.e. the interaction of heavy quarks with the hot
medium. The comparison of prompt D-meson and non-prompt J/$\psi$
$R_{\rm AA}$ shows an indication of larger suppression for D
mesons with respect to B mesons, that can confirm the mass dependent
nature of the energy loss mechanisms.







\end{document}